\documentclass[twocolumn,10pt,twoside]{IEEEtran}
\usepackage{amsmath}
\usepackage{amsthm,amssymb,amsmath,bm}
\usepackage{subfigure}
\usepackage{amsfonts}
\usepackage{subfigure}
\usepackage{amsmath}
\usepackage{epsfig}
\usepackage{amssymb}
\usepackage{amsmath}
\usepackage{cite}
\usepackage[utf8x]{inputenc}
\usepackage{tcolorbox}
\usepackage{multirow}
\usepackage{rotating}
\usepackage{graphicx}
\usepackage{tabularx}
\usepackage{array}
\usepackage{color,soul}
\usepackage{bm}
\usepackage{tikz}
\usepackage{graphicx,dblfloatfix}
\usepackage{blindtext}

\newtheorem{theorem}{Theorem}

\newtheorem{remark}{Remark}

\DeclareMathOperator*{\argmax}{arg\,max} 

\newcommand\minXc[1]{\left[#1\right]^{N} }
\usepackage{subfigure}
\usepackage{moreverb}
\usepackage{epsfig}
\usepackage{amsmath,amssymb,amsthm,mathrsfs,amsfonts,dsfont}
\usepackage{adjustbox,lipsum}
\usepackage{amsfonts}
\usepackage{epsfig}
\usepackage{amssymb}
\usepackage{amsmath}
\usepackage{amsthm}
\usepackage{subfigure}
\usepackage{multirow}
\usepackage{rotating}
\usepackage{graphicx}
\usepackage{tabularx}
\usepackage{array}
\usepackage{anyfontsize}
\usepackage{color,soul}
\usepackage{graphicx,dblfloatfix}
\usepackage{epstopdf}
\usepackage{blindtext}
\usepackage{amsmath}
\usepackage{amsthm,amssymb,amsmath,bm}
\usepackage{subfigure}
\usepackage{amsfonts}
\usepackage{epsfig}
\usepackage{amssymb}
\usepackage{amsfonts}
\usepackage{amsmath}
\usepackage{cite}
\usepackage{graphicx}
\usepackage{fancyhdr}
\usepackage{subfigure}
\usepackage[subfigure]{tocloft}
\usepackage[font={small}]{caption}
\usepackage{subfigure}
\usepackage{tabularx}
\usepackage{tcolorbox}
\allowdisplaybreaks
\usepackage{cite}
\usepackage[linesnumbered,ruled,vlined]{algorithm2e}
\SetKwInput{KwInput}{Input}
\SetKwInput{KwOutput}{Output}
\usepackage{amsthm,amssymb,amsmath,bm}
\usepackage{graphicx}
\usepackage{fancyhdr}
\usepackage{subfigure}
\usepackage[subfigure]{tocloft}
\usepackage[font={small}]{caption}
\usepackage{subfigure}
\usepackage{tabularx}
\usepackage{cite}
\usepackage{url}
\usepackage{hyperref}

\def\black{\textcolor{black}}
\allowdisplaybreaks    
\begin{document}
\title{ 
 Query-Age-Optimal
 Scheduling 
 under Sampling and Transmission Constraints
}
\author{\IEEEauthorblockN{Abolfazl Zakeri, Mohammad Moltafet,  Markus Leinonen, and 
  \IEEEauthorblockN{Marian Codreanu}
  }
  \vspace{-0.5 em}
 \thanks{
 A. Zakeri and M. Leinonen are with 
 CWC-RT,
   University of Oulu, Finland,
 e-mail: \{abolfazl.zakeri, markus.leinonen\}@oulu.fi. 
 M. Moltafet is with Department of Electrical and Computer Engineering University of California Santa Cruz, email: mmoltafe@ucsc.edu.
  M. Codreanu is with Department of Science and Technology,
   Link\"{o}ping University, Sweden, 
   e-mail: marian.codreanu@liu.se.
   This research has been financially supported by the Infotech Oulu, the Academy of Finland (grant 323698), and 6G Flagship program  (grant 346208). The work of M. Leinonen has also been financially supported in part by the Academy of Finland (grant 340171).
 }
\vspace{-2.7 em}
}
	\maketitle
	\begin{abstract} 
	This letter provides query-age-optimal joint sampling and transmission  scheduling policies for a heterogeneous status update system, consisting of a stochastic arrival and a generate-at-will source, with an unreliable channel.
	Our main goal is to minimize the average query age of information (QAoI) subject to average sampling, average transmission, and  per-slot transmission constraints.
	To this end, an optimization problem is formulated and solved by casting it into a linear program. 
We also provide a 
	low-complexity  near-optimal policy 
	using the notion of \textit{weakly-coupled} constrained Markov decision processes.
	The numerical results show up to $32\%$
performance improvement	by the proposed policies compared with a benchmark policy.
\end{abstract}
\vspace{-0.5 mm}
\vspace{-1.6 em}
	\section{Introduction}
The \textit{age of information} (AoI) has been proposed to characterize the information freshness in status update systems
\cite{Roy_2012}. The  AoI is defined as the time elapsed since the latest received status update packet was generated \cite{Roy_2012, AoI_Monograph_Modiano}.
Minimizing the AoI is mainly associated with sampling and 
scheduling optimization. 
However, in some cases, sampling cannot be controlled \cite{Anthony_CSMA_2021},
e.g., a sensor takes a sample whenever it harvests enough energy.
At the same time,
controllable and uncontrollable\footnote{Controllable source refers to the generate-at-will source that generates updates upon request, while uncontrollable source refers to a source that generates updates randomly and cannot be prompted to produce them.}
sources can coexist and utilize shared resources.
 \black{Besides, sampling and transmission of a status update incur different costs involving \textit{distinct non-transferable} resources, leading to individual trade-offs between each resource usage and freshness.
 These motivate us to address
 a freshness optimization problem in a
 \textit{heterogeneous} status update system, consisting of controllable and  uncontrollable sources, 
 \black{subject to \textit{individual} sampling and transmission constraints.}
 }
\\\indent
We consider two sources,
one generate-at-will and one stochastic arrival,
with a transmitter communicating with a remote monitor over an error-prone channel (see Fig. \ref{SM}). 
The transmitter is in charge to sample  and transmit updates to the monitor.
The freshness at the monitor is captured by 
a generalized form of the AoI, the \textit{query AoI} \cite{Petar_QAoI_J} (QAoI).

We formulate an optimization 
problem aiming to minimize the sum (time) average QAoI at the monitor subject to average sampling, average transmission, and  per-slot transmission constraints. 
We then develop optimal (joint sampling and scheduling) 
polices 
by solving the 
problem
via casting it as a constrained Markov decision process (CMDP) problem and subsequently, as a linear program (LP). To strike a  balance between complexity and performance, a near-optimal low-complexity policy is also developed.
The following summarizes  the main contributions of the letter:
\\
$\bullet$ We provide 
    a query-age-optimal 
       policy for a heterogeneous status update system   under average
 sampling and transmission constraints. 
 To this end, we use the LP approach to 
solve 
    a \textit{multi-constraint} CMDP problem. 
    \\ $\bullet$ We devise a \textit{near-optimal low-complexity}  policy. 
To do so,  
we cast and solve a \textit{weakly coupled} CMDP problem \cite{WC_CMDP_Mukul} and  propose a dynamic truncation algorithm.
\\ $\bullet$ Finally, simulation evaluations are conducted
that verify  the effectiveness of the proposed policies.

\vspace{-0.4 mm}
\black{\textit{Related work:}
Recently, the information 
freshness has gained much attention from the sampling and/or scheduling optimization perspectives by using the AoI metric in, e.g.,  \cite{ 
Jnt_Smpl_NonSlot_Sched_May2022, Elif_ARQ, Marian_Scheduling_Cost, Shroff_sampling_2022,Yin_Sun_2, Eytan_Sch2,  Walid_Sam_Updat1, JSAC_2022}, and using the query/request-based AoI metric in, e.g., \cite{Waiting_But_not_Agin_ACM2021,Petar_QAoI_J, Elif_QAoI_2021}.
Assuming the generate-at-will model, the sampling times in \cite{Jnt_Smpl_NonSlot_Sched_May2022,  Marian_Scheduling_Cost, Elif_ARQ, Shroff_sampling_2022,  Eytan_Sch2, Yin_Sun_2, Walid_Sam_Updat1, Petar_QAoI_J, Elif_QAoI_2021}, and the sampling rate in \cite{Jnt_Smpl_NonSlot_Sched_May2022} were optimized.
The work \cite{Shroff_sampling_2022}  considered a limit on the sampling frequency. The works \cite{Elif_ARQ,  Petar_QAoI_J}   considered the transmission cost, which limits the average number of transmissions \cite{Elif_ARQ}.  
\textcolor{black}{
Nevertheless, in some applications,
 sampling of the monitored process and transmission of an update \textit{both}  incur significant, yet different costs \cite{Walid_Sam_Updat1}.
However, only a few works \cite{Jnt_Smpl_NonSlot_Sched_May2022, Marian_Scheduling_Cost, Walid_Sam_Updat1, JSAC_2022} studied \textit{joint} sampling and scheduling control with sampling and transmission costs. These models incorporated a \textit{weighted sum} of the costs.
In contrast, we incorporate the costs separately and consider two individual constraints,  one on the average sampling cost and one on the average transmission cost. 
This is motivated by the fact that generally, the sampling  and transmission costs might need to be imposed on the design by distinct and non-transferable budgets, i.e., the budgets on different types of resources (e.g., CPU and spectrum), so that 
savings in 
one budget do not create a surplus to be used in the other one.  
Consequently, casting the underlying optimization problem as a CMDP inevitably leads to a multi-constraint CMDP problem. 
Solving such a problem 
is in general more challenging than solving the most studied single-constraint CMDP problems in, e.g.,  \cite{Elif_ARQ, Yin_Sun_2, Jnt_Smpl_NonSlot_Sched_May2022, Walid_Sam_Updat1}.
}
}

\textcolor{black}{
In summary,
the main distinctive contribution of this letter is to consider distinct and non-transferable sampling cost and transmission cost constraints,
whereas the prior works have considered a single constraint to account for the both costs. 
To that extent, our provided solutions can be applied to multi-constraint CMDP problems, whereas most prior works are limited to a single-constraint CMDP problem.
}

\vspace{-1. em }
	\section{System Model and Problem Formulation}
\begin{figure}[t!]
    \centering
    \includegraphics[width=.4\textwidth]{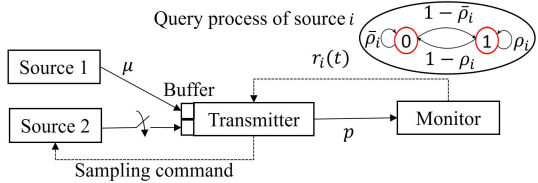}
    \vspace{-3mm}
    \caption{
The considered heterogeneous status update system. 
}
    \label{SM}
    \vspace{- 2 em }
\end{figure}

We consider a heterogeneous status update system consisting of two independent sources: 
    one random arrival (Source 1)  and one generate-at-will (Source 2), a buffer-aided transmitter, and a monitor, as depicted in  Fig. \ref{SM}.
	A discrete-time system with unit time slots $t \in \{0, 1,  \dots\}$ is considered.
		 	The sources are indexed
by $i \in \{1, 2\}$.
The status update packets of Source 1 are generated according to the Bernoulli  distribution with parameter $\mu$.
\textcolor{black}{	
 To maintain the freshest information currently available at the transmitter,
	the most recently arrived update of Source 1 and the most recently sampled update of Source 2 are stored in the transmitter's buffers of size one packet per source.}
	Notice that considering a buffer of size one packet for each source is sufficient in our system, because storing and transmitting the outdated packets does not improve the AoI.
\begin{remark}
    \textcolor{black}{Two sources are considered for the clarity of the  presentation.
	However,	the
 		 formulated problem and solutions  can be extended to  an arbitrary number of sources.
    The extension can be obtained by
    introducing a set of random arrival sources $\mathcal{L}$ and  a set of generate-at-will sources $\mathcal{K}$, then replacing $i=1$ by $l,\,\forall\,l\in\mathcal{L}$ and $i=2$ by $k,\,\forall\,k\in\mathcal{K}$ in the formulas. 
    We will examine the impact of the number of sources on the system performance in  Sec. \ref{Sec_Numerical_Res} (see Fig. \ref{Fig_NS}).}
\end{remark}
\indent
To capture the randomness of the wireless link between the transmitter and the monitor, we assume the link is error-prone, i.e., the reception of updates  by
the monitor are successful
with a constant (over slots) probability ${ p\in(0, 1]}$.
Moreover, we assume that at each slot, at most one packet can be transmitted over the link, which stands for the \textit{per-slot transmission constraint} in the system. 
Besides, we assume that each transmission occupies one slot, and perfect feedback (i.e., instantaneous and error-free) is
available for the link. 
\\\indent
\textit{Decision Variables:}
We define two decision variables: one for the scheduling at the transmitter and the other for the sampling decision  for Source 2. Let $\alpha_i(t) \in \{0,1\}$, $i\in \{ 1 , 2\} $, denote the transmission
decision of source $i$ at the transmitter in slot $t$, where $\alpha_i(t) = 1$ means that the
transmitter sends the packet of source $i$, and $\alpha_i(t) = 0$ otherwise. 
Moreover, let $\beta(t)\in\{0,1\}$ denote the sampling decision of Source 2: $\beta(t)=1$ indicates that the transmitter takes a new sample of Source 2 in slot $t$, and $\beta(t)=0$ otherwise. 
\\\indent
\textit{Age of Information:}
First, we make a common assumption (see e.g., \cite{Walid_Sam_Updat1, Marian_Scheduling_Cost} and  references therein)   that the AoI values
are upper-bounded by a sufficiently large value $N$. 
Besides tractability, this accounts for the fact that when the status information
becomes excessively stale by reaching $N$, the time-critical end application would not be affected if counting further.
Let $\theta_i(t)$  be the AoI of source $i$ at the transmitter in slot $t$, and $\delta_i(t)$ be the AoI of source $i$ at the monitor in slot $t$.
Let
$q_i(t) \in\{0,1\},\,i\in\{1,2\}, $ be an indicator  for  source $i$, where $q_i(t) = 1$ means that the packet of source $i$ is transmitted and successfully received in  slot $t$, and ${q_i(t) = 0}$ otherwise.

Assuming that the packet arrivals occur at the beginning of slots,
the evolution of the ages for $t=1,2,\dots$, with the initial values $\theta_i(0) = 0$ and $\delta_i(0)=0$, are given as  
	\begin{equation}
\begin{array}{ll}\label{Eq_AoIDy}
  &  \theta_1(t+1)= \left\{\begin{array}{ll}
     0,  & \text{if a packet arrives in $t+1$}, \\ 
        \minXc{\theta_1(t)+1}, & \text{otherwise},  
  \end{array} \right.
    \\&
    \theta_2(t+1)=\left\{\begin{array}{ll}
    0,  & \text{if}~~ \beta(t+1)=1, \\
         \minXc{\theta_2(t)+1}, & \text{otherwise}.
     \end{array} \right.
\\ & 
\delta_1(t+1)=\left\{\begin{array}{ll}
    \minXc{\theta_1(t)+1}, & ~\text{if}~
   q_1(t) = 1,
    \\
    \minXc{\delta_1(t)+1}, &
    ~\text{otherwise},
    \end{array} \right.
    \\ & 
\delta_2(t+1)=\left\{\begin{array}{ll}
    1, & \text{if}~
     q_2(t) = 1 , \beta(t) = 1,
    \\
         \minXc{\theta_2(t)+1}, & \text{if}~
   q_2(t) = 1 , \beta(t) \neq 1,
    \\
   \minXc{\delta_2(t)+1 }, &\text{otherwise},
    \end{array} \right.
    \end{array}
\end{equation}
where ${\left[.\right]^N } \triangleq \min(.\,, N)$.
\\\indent
We consider that the AoI at the monitor becomes important (only) in the instants when the updates are actually needed/used by the time-critical end application. 
We capture this by the QAoI \cite{Petar_QAoI_J} by defining 
a per-source query flag denoted  by $ { r_i(t)\in \{0,1\},\,i\in\{1,2\} },$ where $r_i(t) = 1$ means that the monitor queries the information of source $i$ in slot $t$, and $r_i(t) = 0$ otherwise. 
Accordingly, the QAoI for source $i$ in slot $t$ is defined as $r_i(t) \delta_i(t)$.
We assume that the query flags are generated according to the binary Markov chains, as depicted in Fig. \ref{SM}, where $\rho_i$ and $\bar{\rho}_i$ are, respectively, the self-transition probabilities of State $1$ and State $0$.

We denote the sum  (discounted) average QAoI
      at the monitor by $\bar{\delta}$, defined by 
 \begin{equation}
     \begin{array}{ll}
    & \displaystyle \bar{\delta}\triangleq
        \textstyle
        \bar{\lambda}
        \sum_{t=1}^{\infty} 
      {\lambda^{t-1}}
         \mathbb{E}\left\{\sum_i r_i(t)\delta_i(t)
        \right\} ,
      \end{array}
 \end{equation}
 where $\lambda\in(0,1)$ is the discount factor, \black{$\bar{\lambda}=(1-\lambda)$ is the normalization factor \cite[Remark 2.1]{Eitam_CMDP}}, and $\mathbb{E}\{\cdot\}$ is the expectation with respect to the random channel, the packet arrival process, and the (possibly random) decision variables $\alpha_i(t)$ and $\beta(t)$ made in reaction to the past and current AoI  and query values.
 We also define the (discounted) average number of transmissions, denoted by $\bar{\alpha}$, and   the (discounted) average number of sampling actions, denoted by $\bar{\beta}$, as
   \begin{align}
   \nonumber
    \displaystyle \bar{\alpha} \triangleq 
    \bar{\lambda}
       \textstyle
        \sum_{t=1}^{\infty} 
      \lambda^{t-1}
         \mathbb{E}\left\{\sum_i \alpha_i(t)
        \right\},~
        \displaystyle \bar{\beta} \triangleq
        \bar{\lambda}
        \textstyle
        \sum_{t=1}^{\infty}
         \lambda^{t-1}
         \mathbb{E}\left\{\beta(t)
        \right\}.
 \end{align}
Our main goal is to 
solve the following problem\footnote{
     As proposing the CMDP approach to the main problem \eqref{Org_P1},
     the provided solutions can be extended to the average (expected) case (i.e., $\lambda = 1$) if the unichain structure \cite[Sec. 8.3.1]{Puterman_Book} exits.
   \black{The unichain structure, however, does not exist for our problem since one can construct a deterministic policy which induces a Markov chain with two recurrent classes. Thus,  for the average case, finding a solution could be, in general, difficult \cite[Sec. 9.2.6]{Kallenberg_MDPBook}.}}:
\allowdisplaybreaks
\vspace{- .5 em}
        \begin{subequations}
       \label{Org_P1}
       \begin{align}
        		\underset{\{\{\alpha_i(t)\}_{i=1}^{2},\beta(t)\}_{t=1,2,\ldots}}
          {\mbox{minimize}}~~~   &
        		\bar{\delta}
        		\vspace{-2em}
        		\\
        		\mbox{subject to}~~~ & 
        	 \bar{\alpha}
       \le \Gamma^{\mathrm{tr}}, 
                   \label{Con_Tra1}
                \\&
                 \bar{\beta}
       \le \Gamma^{\mathrm{sm}}, 
                   \label{Con_Sam1}
                   \\& 
                   \label{Cons_capacity_1}
                   \textstyle\sum_i \alpha_i(t)\le 1,~\forall\,t,
                   \\& 
                   \label{Cons_smp_trans_S2}
                 \beta(t) \le \alpha_2(t), ~\forall\,t,
                \vspace{-1.2 em}
                   \end{align}
        		\end{subequations} 
        	 with the decision variables ${ \{\{\alpha_i(t)\}_{i=1}^{2},\beta(t)\}_{t=1,2,\ldots} }$, 
        		where   ${ \Gamma^{\mathrm{tr}}\in(0,1] }$ and $ { \Gamma^{\mathrm{sm}} \in(0,1] } $  
        		are limits on the 
        		 average 
        		number of transmissions and sampling actions in the system.
	The per-slot constraint \eqref{Cons_capacity_1} limits that at most one packet can be transmitted over the link in each slot. The per-slot constraint \eqref{Cons_smp_trans_S2} ensures  the sampled update in slot $t$ is transmitted at the same slot.
Note that
	 constraint \eqref{Cons_smp_trans_S2} is not obligatory for  problem \eqref{Org_P1}, but
 it 	eliminates a sub-optimal action, i.e., taking a new sample without its concurrent transmission, 
	which simplifies solving the CMDP problem (introduced next). 
	
	 Next, we   provide an optimal policy to problem \eqref{Org_P1}.

\vspace{-2 mm}
\section{ Optimal Policy }
To solve  problem \eqref{Org_P1}, we first recast it into a CMDP problem and then (optimally) solve it via
solving its equivalent LP  \cite{Eitam_CMDP}. 
\vspace{-5 mm }
\subsection{CMDP Formulation of Problem \eqref{Org_P1}}\label{Sec_CMDP}
The CMDP is specified  by the following elements:
\\$\bullet $
    \textit{State}:
    We define the state in slot $t$ by
 ${ \bold{s}(t)\triangleq \big (r_1(t),\theta_1(t),\delta_1(t),r_2(t),\theta_2(t),\delta_2(t)\big ) }$.
   We denote the state space  by  $\mathcal{S}$ which is a finite set.
     \\$\bullet $ \textit{Action}:   
     \textcolor{black}{  
    Actions determine the transmission and sampling decisions while incorporating  constraints \eqref{Cons_capacity_1} and \eqref{Cons_smp_trans_S2}.
     Thus, we define the action taken in slot $t$ by $ { a(t)\in\{0,1,2_\mathrm{r},2_\mathrm{n}\} }$, where ${a(t)=0}$ means  that the transmitter stays idle, ${a(t)=1}$ means
     that the packet of Source 1 is  transmitted/re-transmitted,
     ${a(t)=2_\mathrm{r}}$ means that the packet of Source 2 is  re-transmitted, and ${a(t)=2_\mathrm{n}}$ means that  a new sample of Source 2 is transmitted. 
     }
     Also,  $\mathcal{A}$
  is the action space of size $|\mathcal{A}|=4$. 
     Actions are determined by a policy $\pi$, which is a (possibly randomized) mapping from $\mathcal{S}$ to $\mathcal{A}$.
     \\$\bullet $ \textit{State Transition Probabilities}:
     We denote the state transition probability from state $\bold{s}$ to next state $\bold{s}'$ under an action ${a}$ by $\mathcal{P}_{\bold{s}\bold{s}'}(a)$.
    Since the evolution of the AoIs in \eqref{Eq_AoIDy} and the query processes are independent among the sources, the transition probability can be written as
     $ 
  {
     \mathcal{P}_{\bold{s}\bold{s}'}(a)=\prod_{i} 
      \Pr\{r_i'\,|\,r_i\}
     \Pr\{\bold{\underline{s}}_i' 
     \,\big|\,\bold{\underline{s}}_i,a\} }$, 
where 
$\Pr\{r_i'\,|\,r_i\}$ is given by \eqref{Eq_TP_query}, and 
$\Pr\{\bold{\underline{s}}_i'\,\big|\,\bold{\underline{s}}_i,a\}$ 
is given by \eqref{Eq_TranPro_Unr},
in which 
$\bold{\underline{s}}_i \triangleq (\theta_i,\delta_i)$ contains the current age values, as a part of the current state,  associated with source $i$, $\bold{\underline{s}}_i' \triangleq(\theta_i',\delta'_i)$ contains the next age values, as a part of the next state,  associated with source $i$, and
 $\bar{\mu}=1-\mu$.
\allowdisplaybreaks
        \begin{align}
            \label{Eq_TranPro_Unr}
  &   \Pr\{\bold{\underline{s}}_1'\,\big|\,\bold{\underline{s}}_1,a\}=
     \\
  &  
  \nonumber
  \left\{ \begin{array}{ll}
      \mu, & a \neq 1;~ \theta'_1=0,~\delta'_1=\minXc{\delta_1+1},
      \\
     \bar{\mu}, & a \neq 1;~ \theta'_1 = \minXc{\theta_1+1},\, \delta'_1=\minXc{\delta_1+1},  
      \\
  \mu p, & a = 1;~ \theta'_1=0,~\delta'_1=\minXc{\theta_1+1},
  \\
  \mu (1-p), & a = 1;~ \theta'_1=0,~\delta'_1=\minXc{\delta_1+1},
  \\
   \bar{\mu} p, & a = 1;~ \theta'_1=\minXc{\theta_1+1},~\delta'_1=\minXc{\theta_1+1},
  \\
  \bar{\mu}(1-p),
 & a = 1;~ \theta'_1=\minXc{\theta_1+1},~\delta'_1=\minXc{\delta_1+1},
      \\ 0& \text{otherwise}.
      \end{array} \right.
 \\ &   \Pr\{\bold{\underline{s}}_2'~\big|~\bold{\underline{s}}_2,a\}= 
  \nonumber
     \\ &
     \nonumber
   \left\{ \begin{array}{ll}
      1, & a \neq \{2_\mathrm{r},2_\mathrm{n}\};\,
     \theta'_2 = \minXc{\theta_2+1},\, \delta'_2=\minXc{\delta_2+1},
      \\
     p, & a = 2_\mathrm{r};~ \theta'_2= \minXc{\theta_2+1},~\delta'_2=\minXc{\theta_2+1},
      \\
   1-p, & a = 2_\mathrm{r};~ \theta'_2=\minXc{\theta_2+1},~\delta'_2=\minXc{\delta_2+1},
  \\
    p, & a = 2_\mathrm{n};~ \theta'_2=1,~\delta'_2=1, 
      \\
   1-p, & a = 2_\mathrm{n};~ \theta'_2=1,~\delta'_2=\minXc{\delta_2+1},
      \\ 0& \text{otherwise}.
      \end{array} \right.
      \end{align}
         \begin{equation}
    \label{Eq_TP_query}
     \Pr\{r_i'\,|\, r_i\} = 
  \left\{ \begin{array}{ll}
  \rho_i, & r_i = 1;~  r'_i =1,
  \\
  1-\rho_i,  & r_i = 1;~  r'_i =0,
  \\
  \bar{ \rho}_i, & r_i = 0;~  r'_i =0,
  \\
  1- \bar{ \rho}_i,  & r_i = 0;~  r'_i =1.
  \end{array} \right.
  \end{equation}
   $\bullet $ \textit{Cost Functions}:
      The  cost functions include: 1) the QAoI cost defined by $ c(\bold{s}(t))= \sum_i r_i(t) \delta_i(t)  $, 
 2) the transmission cost defined by $d^{\mathrm{tr}}(a(t))=\mathds{1}_{\{a\neq 0\}}$, and 3) the sampling cost defined by $d^{\mathrm{sm}}(a(t))=\mathds{1}_{\{a= 2_\mathrm{n}\}}$, 
where
   $\mathds{1}_{\{\cdot\}}$ is an indicator function  which   equals to $1$ when the condition in $\{\cdot\}$ holds. 
\\\indent
By the above definitions, 
problem \eqref{Org_P1} can  
be  recast as the following CMDP problem
        \begin{subequations}
       \label{Pro_CMDP_1}
       \begin{align}
       \underset{\pi\in\Pi}{\text{minimize}}~ & 
        \bar{\lambda}
     \textstyle  \sum_{t=1}^{\infty}
     \lambda^{t-1}
         \Bbb{E}^{\pi, \eta} \left\{ 
         c(\bold{s}(t))
        \right\},
        		\\
        		\mbox{subject to}~  &
        \label{Eq_Avr_Tra}
      \bar{\lambda}
     \textstyle  \sum_{t=1}^{\infty}
       \lambda^{t-1}
         \Bbb{E}^{\pi, \eta} \left\{ d^{\mathrm{tr}}(a(t))
         \right\} \le \Gamma^{\mathrm{tr}},
                \\&
                \label{Eq_Avr_Smp}
          \bar{\lambda}
     \textstyle  \sum_{t=1}^{\infty}
       \lambda^{t-1}
         \Bbb{E}^{\pi, \eta} \left\{ d^{\mathrm{sm}}(a(t))
         \right\} \le \Gamma^{\mathrm{sm}},
                   \end{align}
        		\end{subequations} 
where $\Pi$ is the set of all stationary randomized policies, and $\Bbb{E}^{\pi, \eta}$ denotes the expectation when following a policy $\pi$ for a given initial distribution (over the state space) $\eta$.
\\\indent
Next, we provide an optimal policy to  problem \eqref{Pro_CMDP_1}.
\vspace{-1 em}
	\subsection{Linear Programming of the CMDP Problem \eqref{Pro_CMDP_1}} \label{Subsec_LP}
Here, we transform problem \eqref{Pro_CMDP_1} into an LP. 
	For a notional simplicity, we denote state $\bold{s}$ by $s$. 
	Let ${ x(s,a)\in\Bbb{R}^{|\mathcal{S}||\mathcal{A}|}}$, ${\forall\,s\in\mathcal{S},~\forall\,a\in\mathcal{A}}$, be defined as \cite[Ch. 3]{Eitam_CMDP}
	\begin{align}
	\nonumber
	\begin{array}{ll}
	    x(s,a) \triangleq \bar{\lambda} \sum_{t=1}^{\infty}
	    \lambda^{t-1}
	    \Pr^{\eta}{\{s(t)=s,\,a(t)=a\}},
	    \end{array}
	\end{align}
	where $\Pr^{\eta}{\{s(t)=s,\,a(t)=a\}}$ denotes the probability of taking action $a$ at state $s$ in slot $t$ given the initial distribution $\eta$.
It is noteworthy that  $x(s,a),~\forall\,s\in\mathcal{S},~\forall\,a\in\mathcal{A},$  can be interpreted as the long
run discounted time that the system is in state $s$ and action
$a$ is chosen.
Then, the CMDP problem \eqref{Pro_CMDP_1} can be transformed into the following LP 
\cite[Ch. 3]{Eitam_CMDP}:
\begin{subequations}\label{Prob_LP_1}
  \begin{align}
     \underset{\bold{X}}{\text{minimize}}~~
      & 
      \textstyle\sum_{s\in\mathcal{S}}\sum_{a\in\mathcal{A}} x(s,a)c(s)
 \\
      \text{subject to}~~ &
      \label{Eq_LP_Targ}
     \textstyle\sum_{s\in\mathcal{S}}\sum_{a\in\mathcal{A}} x(s,a)d^{\mathrm{tr}}(a) \le \Gamma^{\mathrm{tr}},
     \\& \label{Eq_LP_sm}
        \textstyle\sum_{s\in\mathcal{S}}\sum_{a\in\mathcal{A}} x(s,a)d^{\mathrm{sm}}(a) \le \Gamma^{\mathrm{sm}},
        \\& \nonumber
         \textstyle \sum_{s'\in\mathcal{S}} \sum_{a\in\mathcal{A}} x(s',a)\big(\mathds{1}_{\{s=s'\}} - \lambda \mathcal{P}_{s's}(a) \big) 
         \\ & \label{Eq_LP_balance}
         = \bar{\lambda}\eta(s),
         \forall\, s\in\mathcal{S},
         \\
        &  \label{Eq_LP_positive}
        x(s,a)\ge 0,~\forall\, s\in\mathcal{S},~\forall\, a\in\mathcal{A},
            \end{align}
     \end{subequations}
     with variables $\bold{X}=[x(s,a)]$, where $\eta(s),\,\forall\, s\in\mathcal{S}$, is the probability that the initial state is state $s$. 
     By \cite[Theorem 3.3]{Eitam_CMDP}, 
     the following theorem relates a solution of the LP \eqref{Prob_LP_1} to a solution of 
     problem \eqref{Pro_CMDP_1}.
     \begin{theorem}
First, the optimal value of the CMDP problem \eqref{Pro_CMDP_1} equals to the optimal value of the LP \eqref{Prob_LP_1}. 
Moreover, 
let ${ \bold{X}^*=[x^*(s,a)] },$ be a solution of the LP \eqref{Prob_LP_1}. Then, a stationary randomized policy $\pi\triangleq\{f(s,a)\}$ with
\begin{align}
\label{Eq_Optimal_Policy}
  \begin{array}{cc}
 f(s,a)=
  \left\{ 
  \begin{array}{ll}
    x^*(s,a)/\bar{x}_s, & \text{if}~\bar{x}_s>0, \\
   \text{arbitrary},  & \text{otherwise},
       \end{array}\right.
\end{array}  
\end{align}
is an optimal policy to the CMDP problem \eqref{Pro_CMDP_1}, 
where $ { \bar{x}_s=\textstyle\sum_{a'\in\mathcal{A}}x^*(s,a') } $, and $f(s, a),~\forall\, s\in\mathcal{S},~\forall\, a\in\mathcal{A},$ is the probability that an action $a$ is chosen at state  $s$.
     \end{theorem}
\textbf{Complexity analysis:} 
\textcolor{black}{
The complexity of finding the optimal policy (determined by \eqref{Eq_Optimal_Policy}) 
comes from
the complexity of the LP \eqref{Prob_LP_1}, which is exponential in the number of sources. This is because
 its number of variables and constraints vary with the \textit{state space size} of the CMDP problem \eqref{Pro_CMDP_1}, which
grows exponentially with the number of sources.}
Thus, the optimal policy becomes  computationally inefficient when applied for large numbers of sources. 
To strike a balance between the performance and complexity, a heuristic  low-complexity policy is provided in the next section.
	\section{Low-Complexity  Policy }
	In this section, we develop a low-complexity policy to the main problem \eqref{Org_P1} via the following procedure: 1) we relax the per-slot transmission constraint \eqref{Cons_capacity_1}, 2)  transform the relaxed problem to a weakly coupled CMDP problem which is then solved by its equivalent LP
	\cite{WC_CMDP_Mukul}, and 3) propose a dynamic truncation algorithm to satisfy the   transmission  constraint \eqref{Cons_capacity_1}. 
	\\\indent
We relax the per-slot  constraint \eqref{Cons_capacity_1} to the (discounted) time average constraint 
$ \bar{\lambda} \sum_{t=1}^{\infty}
	    \lambda^{t-1}\mathbb{E}\{\textstyle\sum_i \alpha_i(t)\}\le 1 $.
This constraint then becomes inactive due to  constraint \eqref{Con_Tra1}. 
Then, the main  problem \eqref{Org_P1} without  constraint \eqref{Cons_capacity_1}
can be cast as a weakly coupled CMDP problem.
This is because the transition probabilities, the action space, and the immediate  cost  functions  become independent among the sources; \black{now, the average transmission constraint \eqref{Con_Tra1} is the only constraint coupling the sources.} 
\\\indent
To proceed, we define the state, action, transition probabilities, and cost functions associated with each source $i$. The state of source $i$ in slot $t$ is ${ s_i(t) \triangleq \big( r_i(t), \theta_i(t),\delta_i(t) \big) }$, with state space $\mathcal{S}_i$, which is a finite set.
The action of source $i $ in slot $t$ is denoted by $a_i(t)\in\mathcal{A}_i$ with action spaces $\mathcal{A}_1=\{0,1\}$ and $\mathcal{A}_2=\{0, 2_\mathrm{r}, 2_\mathrm{n}\}$.
The state transition probabilities of source $i$, denoted 
as ${ \mathcal{\tilde{P}}_{s_is_i'}^{i}(a),~\forall\, s_i,\,s_i'\in\mathcal{S}_i }$,
is given by
     $ \Pr\{r_i' \,|\,r_i \}
     \Pr\{\bold{\underline{s}}_i' 
     \,\big|\,\bold{\underline{s}}_i,a\}
$, where        $\Pr\{r_i' \,|\,r_i\}$ and $
     \Pr\{\bold{\underline{s}}_i' 
     \,\big|\,\bold{\underline{s}}_i,a\}$  were defined in
\eqref{Eq_TP_query} and \eqref{Eq_TranPro_Unr}. The QAoI cost of source $i$ is ${ c_i({s}_i(t))=r_i(t)\delta_i(t) }$. 
The transmission cost of source $i$ is  ${ d_i^{\mathrm{tr}}(a_i(t))=\mathds{1}_{\{a_i(t)\neq 0 \}} }$. The sampling cost of Source 2 is  $ { d^{\mathrm{sm}}(a_2(t))=\mathds{1}_{\{a_2(t)= 2_n \}} }$.
\\\indent
First, for a presentation simplicity, we remove the subscript $i$ from $s_i(t)$ and $a_i(t)$.
Then, for each source $i\in\{1,2\}$, let $x_i(s,a)\in\Bbb{R}^{|\mathcal{S}_i||\mathcal{A}_i|}, ~\forall\,s\in\mathcal{S}_i,~\forall\,a\in\mathcal{A}_i,$ be defined as
	\begin{align}
	\nonumber
	\begin{array}{ll}
	    x_i(s,a)  & \triangleq \bar{\lambda} \sum_{t=1}^{\infty}
	    \lambda^{t-1}
	    \Pr_i^{\eta_i}{\{s(t)=s,\,a(t)=a\}},
	    \nonumber
	    \end{array}
	\end{align}
 	where $\Pr_i^{\eta_i}{\{s(t)=s,\,a(t)=a\}}$ denotes the probability of taking action $a\in\mathcal{A}_i$ at state $s\in\mathcal{S}_i$ in slot $t$ given the initial distribution $\eta_i$.
Then, the corresponding LP of the weakly coupled CMDP problem is   given by 
\begin{subequations}\label{Prob_LP_2}
  \begin{align}
     \underset{\bold{X}_i}{\mbox{minimize}}~~
      &\textstyle\sum_i  \textstyle\sum_{s\in\mathcal{S}_i}\sum_{a\in\mathcal{A}_i} x_i(s,a)c_i(s)
 \\
      \mbox{subject to}~~ &
   \textstyle\sum_i  \textstyle\sum_{s\in\mathcal{S}_i}\sum_{a\in\mathcal{A}_i} x_i(s,a)d_i^{\mathrm{tr}}(a) \le \Gamma^{\mathrm{tr}},
     \\& 
        \textstyle\sum_{s\in\mathcal{S}_2}\sum_{a\in\mathcal{A}_2} x_2(s,a)d^{\mathrm{sm}}(a) \le \Gamma^{\mathrm{sm}},
        \\& \nonumber
         \textstyle \sum_{s'\in\mathcal{S}_i} \sum_{a\in\mathcal{A}_i} x_i(s',a)\big(\mathds{1}_{\{s=s'\}} - \lambda \mathcal{\tilde{P}}^i_{s's}(a) \big)
         \\  &
         = \bar{\lambda} \eta_i(s),
          \forall\,i,~\forall\, s\in\mathcal{S}_i,
         \\
        &   x_i(s,a)\ge 0,\forall\,i,~\forall\, s\in\mathcal{S}_i,~\forall\, a\in\mathcal{A}_i,
            \end{align}
     \end{subequations}
     with variables $\bold{X}_i=[x_i(s,a)],~i\in\{1,2\}$.
     
Let $\bold{X}_i^*=[x_i^*(s,a)]$ be a solution of the LP \eqref{Prob_LP_2}. 
By results of \cite{WC_CMDP_Mukul}, 
an optimal  stationary randomized policy for each source $i$, $\pi^*_i\triangleq\{f_i(s,a)\}$, is given by
\begin{equation}
\label{Eq_Dec_Optm_polcy}
  \begin{array}{cc}
 f_i(s,a)=
  \left\{ 
  \begin{array}{ll}
    x_i^*(s,a)/\bar{x}_{i,s}, & \text{if}~\bar{x}_{i,s}>0, \\
   \text{arbitrary},  & \text{otherwise},
       \end{array}\right.
\end{array}  
\end{equation}
where ${ \bar{x}_{i,s}=\textstyle\sum_{a'\in\mathcal{A}_i}x_i^*(s,a')}$ and  ${ f_i(s, a)}$, ${\forall\, s\in\mathcal{S}_i}$, ${\forall\, a\in\mathcal{A}_i}$, is the probability that an action $a\in\mathcal{A}_i$ is chosen at state  $s\in\mathcal{S}_i$ for source $i$.
\textcolor{black}{Notably, the per-source policies $\{\pi_i^{*}\}$ construct
 a \textit{lower-bound} policy to the main problem \eqref{Org_P1};
 this lower bound is tight as empirically shown in Fig. \ref{Fig_NS}.
}
     \\\indent
     Having constructed the above weakly coupled CMDP and derived its solution,
     we now propose a near-optimal low-complexity policy for the main problem \eqref{Org_P1} 
     that operates as follows:
   (i) it determines an action of each source $i$ in each slot $t$ according to $\pi^*_{i}$, and
   (ii)  having the actions determined, if there is more than one packet to be transmitted, 
  the packet of source $\underline{i}$ will be  transmitted, where  
   ${ \underline{i}=\argmax_{i\in\{1,2\}}\{h_i(t)\} }$ in which 
   $h_1(t)$ and $h_2(t)$ are given as
   \begin{align}
   \label{Eq_h_Slec_Rul}
       \begin{array}{cc}
  h_1(t) \triangleq r_1(t) (\delta_1(t)-\theta_1(t)),~h_2(t) \triangleq  r_2(t)\delta_2 (t).
       \end{array}
   \end{align}
   \indent
    \black{We term the above Step (ii)  as a \textit{dynamic truncation algorithm}, which still couples the sources.}
    The main idea of the truncation algorithm is to transmit the packet that is more fresh from the monitor's perspective, as it has greater potential to decrease the AoI at the monitor.
    To capture such relative freshness, we have used the AoI difference between the monitor and the transmitter for Source 1, and the AoI at the monitor itself for Source 2. 

\textbf{Complexity analysis:} 
  \textcolor{black}{
   The main complexity of finding the low-complexity policy comes from solving the LP
   \eqref{Prob_LP_2}, which is (only) \textit{linear} in the number of sources.
   This is because
    its number of variables and constraints  vary with the \textit{state space size} of the weakly-coupled CMDP problem, which
grows linearly with the number of sources.
}
   Thus, the proposed policy has low complexity,  
   and as numerically shown in the next section,  obtains 
near-optimal performance.
\section{Numerical Results}\label{Sec_Numerical_Res}
\begin{figure}[h]
    \centering
    \includegraphics[width=.35\textwidth]{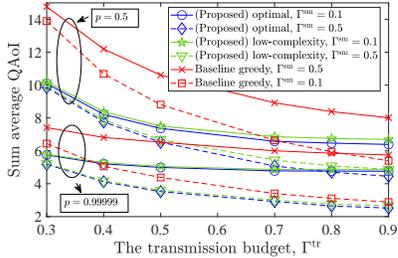}
    \vspace{-.5 em}
    \caption{ \textcolor{black}{
    The sum (discounted) average QAoI versus the transmission budget $\Gamma^{\mathrm{tr}} $ for different channel reliabilities $p$, where $\mu=0.6$.
    }
    \label{Fig_tranBudget}
    }
    \vspace{- 2 em }
    \end{figure}

\begin{figure}[h]
    \centering
    \includegraphics[width=.38\textwidth]{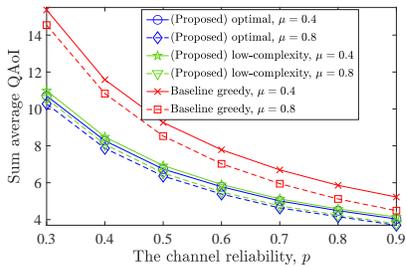}
    \vspace{-0.5 em}
    \caption{ The sum (discounted) average QAoI versus the channel reliability $p$, where $\Gamma^{\mathrm{tr}} = 0.5$ and $\Gamma^{\mathrm{sm}} = 0.3$.
    }
    \label{Fig_chnreliab}
    \vspace{-1.5 em }
    \end{figure}

Here, we assess the performance  of the proposed policies. We set $N = 20$,  $\rho_1=\rho_2=0.7$, $\bar{\rho}_1=\bar{\rho}_2=0.4$, and $\lambda = 0.99999$, and
initialize the query flags by the steady-state probabilities of the Markov chains, unless otherwise stated.
For a benchmark, we consider a greedy-based baseline policy with the decision rule: 
\textit{If $\bar{D}^{\mathrm{tr}}(t)\le \Gamma^{\mathrm{tr}}$, find ${ \underline{i}=\argmax_{i\in\{1,2\}}\{h_i(t)\} }$; otherwise, $a(t) =0$. If $\underline{i} = 1$, then $a(t) = 1$; otherwise, do the following. If $\bar{D}^{\mathrm{sm}}(t) \le \Gamma^{\mathrm{sm}}$, then $a(t) = 2_{\mathrm{n}}$; otherwise, $a(t) = 2_{\mathrm{r}}$}.
     Here, $\bar{D}^{\mathrm{tr}}(t)$ and $\bar{D}^{\mathrm{sm}}(t)$ denote, respectively,  the discounted average number of transmissions and sampling actions until slot $t$, and $h_i(t)$ was given by \eqref{Eq_h_Slec_Rul}.
\\\indent 
Fig. \ref{Fig_tranBudget} shows the sum (discounted) average QAoI performance of the proposed policies 
and the baseline policy as a function of  the transmission budget $\Gamma^{\mathrm{tr}}$ (i.e., the limit on the average number of transmissions). First, Fig. \ref{Fig_tranBudget} (as well as Fig. \ref{Fig_chnreliab}) reveals that the  heuristic policy has near-optimal performance, i.e., it nearly coincides with the optimal policy.
The figure shows the importance of an optimal trade-off design between freshness and resource usage, as the performance gap between the baseline policy and the proposed policies is significant when the transmission budget is small.
Moreover, the figure reveals that the QAoI is more sensitive to the transmission budget than the sampling budget, as expected. 
\\\indent
Fig. \ref{Fig_chnreliab} examines the impact
of the channel reliability $p$ on the sum average QAoI performance of the
different policies for different packet arrival rate  $\mu$ for Source 1. 
 First, the figure shows that the  performance gap between the proposed policies 
and  the baseline policy is large, especially when the channel reliability is small.
The figure also shows that the QAoI considerably decreases as the channel reliability increases because it increases the chance that the monitor receives updates more frequently. 
Moreover, it can be seen that the QAoI would not change much by increasing the arrival rate under a good channel condition. 

   \begin{figure}[h]
    \centering
    \includegraphics[width=.34\textwidth]{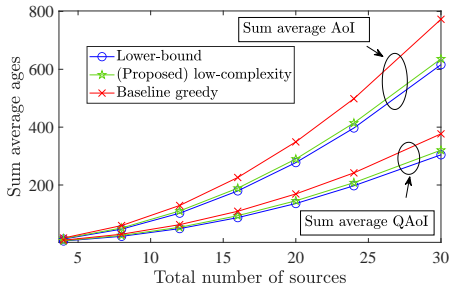}
    \vspace{-0.5 em}
    \caption{ 
    \textcolor{black}{
    The sum (discounted) average AoI and QAoI versus the total number of sources with equal number of random arrival and generate-at-will sources, where 
    $N=150$, $\Gamma^{\mathrm{tr}} = 0.8$, $\Gamma^{\mathrm{sm}} = 0.3$, and $p=0.99999$. Moreover,  $\mu_i=0.5$ (for all the random arrival sources) and
    $\rho_i=\bar{\rho}_i = 0.7$ (for all sources).
    }
    }
    \label{Fig_NS}
    \vspace{-0.5 em }
    \end{figure}
\textcolor{black}{
Fig. \ref{Fig_NS} shows the impact of the 
number of sources 
on the sum average QAoI/AoI performance of different policies. First, the figure shows that the proposed low-complexity policy indeed preserves near-optimal performance in a general multi-source setup, as it nearly coincides with the lower-bound policy. Moreover, the performance gap between the baseline policy and the proposed policy increases in the number of sources. 
}
\section{Conclusion}
We provided a query-age-optimal joint  sampling and scheduling policy for a heterogeneous status update system under  average sampling, average transmission,  and per-slot transmission constraints.
To this end, a CMDP problem with two average constraints was cast and solved via its equivalent LP. 
We also provided a low-complexity policy by casting a weakly-coupled CMDP problem and solving it via its equivalent LP.
Numerical results showed the effectiveness of
the proposed policies, revealing that  age-optimal sampling and scheduling is crucial for
resource-constrained status update systems, where
a greedy-based policy is inefficient.
Moreover, the near-optimal performance of the proposed low-complexity policy substantiates that the LP approach provides a viable solution for practical systems having many sources/sensors.
\bibliographystyle{ieeetr}

\bibliography{Bib/conf_short,Bib/IEEEabrv,Bib/Ref_2SRAoI}
\small
\end{document}